\documentclass[twocolumn,twoside]{revtex4}
\usepackage{graphicx}
\usepackage{fancyhdr}
\begin{document}

\title{Rare B decays and Lepton Flavour Non Universality tests at LHCb}

%

\author{S.~Ferreres Solé\\
on behalf of the LHCb collaboration}
\affiliation{Nikhef, Science Park 105, 1098 XG Amsterdam, The Netherlands}

\begin{abstract}
Summary of some of the most recent measurements performed by LHCb on rare decays and lepton flavour universality tests. 
\end{abstract}

\maketitle

\thispagestyle{fancy}

\section{Introduction}
The analysis of decays hadrons containing charm of beauty quarks increases our understanding of the fundamental constituents of matter and their interactions. Although the Standard Model (SM) has been proven to predict the interaction among particles with great accuracy, there are still some unanswered questions for which physics beyond the Standard Model (BSM) may come into play.

\section{Rare Decays}
One way to search for New Physics is through the analysis of electroweak decays with low branching fractions or even forbidden in the SM. These decays, also known as \textit{rare decays}, might receive contributions from new particles, such as dark matter candidates, altering their branching fractions with respect to the SM predictions. The goal then is to measure the branching fraction of such decays to check if they deviate from the SM predictions.

Two of such rare processes are the $B_{s}^0\rightarrow \mu^+\mu^-$ and $B^0\rightarrow \mu^+\mu^-$ decays, which are of special interest in probing the SM. In the computation of their branching fraction, the decay amplitude can be factorized into the hadronic and leptonic parts, allowing for a clean theoretical calculation of their branching fraction. Because the two final-state particles are muons, their signature in the detector is also very clean. These properties make the $B_{s}^0\rightarrow \mu^+\mu^-$ and $B^0\rightarrow \mu^+\mu^-$ processes very sensitive to BSM physics. 

\begin{figure}[h]
       \centering
       \includegraphics[width=80mm]{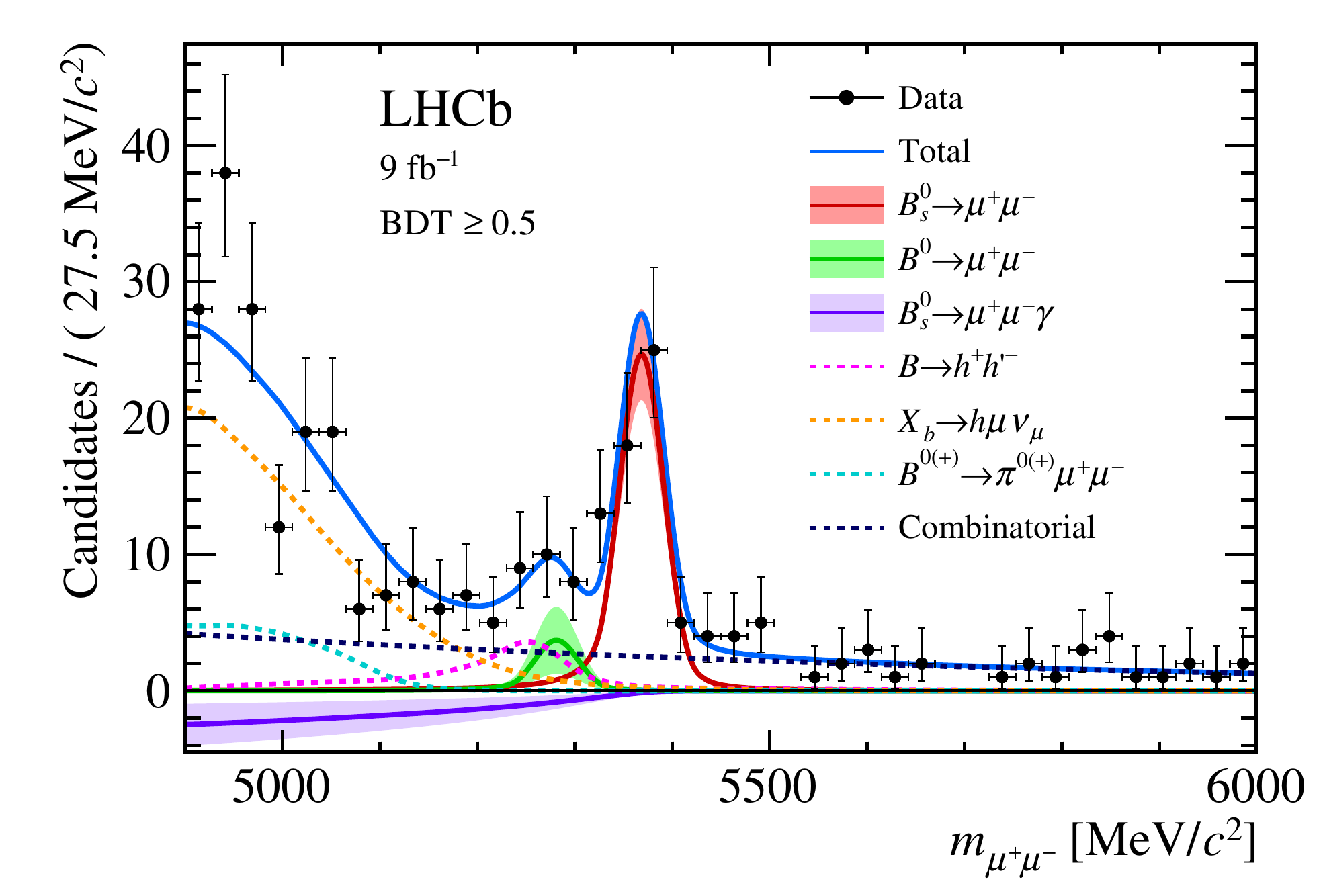}
       \caption{Mass distribution of selected $B_{(s)}^0\rightarrow \mu^+\mu^-$ candidates with BDT $>$ 0.5. The total model and the fit components are ovelaid.} 
       \label{fig:bsmumu}
\end{figure}

The branching fractions of the $B_{s}^0\rightarrow \mu^+\mu^-$ and $B^0\rightarrow \mu^+\mu^-$ decays have been measured by LHCb using full Run 1 and Run 2 data~\cite{prd-bsmumu,prl-bsmumu}, corresponding to a total integrated luminosity of 9 fb$^{-1}$. A pair of opposite charge muons forming a good quality and displaced vertex from the interaction region are selected as the signal candidates. The branching fraction for each decay is determined from maximum likelihood fits to the dimuon invariant mass. The fit is performed simultaneously in bins of a BDT classifier to increase the sensitivity of the data sample. The mass distribution of selected candidates together with the fit components are shown for BDT $>$ 0.5 in Figure~\ref{fig:bsmumu}. From the fit, a branching fraction measurement for $B_{s}^0\rightarrow \mu^+\mu^-$ of $(3.09^{+0.46+0.15}_{-0.43-0.11})\times 10^{-9}$ is obtained. For $B^0\rightarrow \mu^+\mu^-$ no significance excess is found and the upper limit $\mathcal{B}(B^0\rightarrow \mu^+\mu^-) < 2.6\times 10^{-10}$ at 95\% confidence level is set.

In the same analysis, a search for $B_{s}^0\rightarrow \mu^+\mu^-\gamma$ decays considering only initial state radiation and $m_{\mu^+\mu^-} > 4.9$ GeV/c$^2$ has also been performed. Since no significant excess is found, the upper limit $\mathcal{B}(B_{s}^0\rightarrow \mu^+\mu^-\gamma) < 2.0\times 10^{-9}$ at 95\% confidence level is obtained.

Another interesting property of $B_{s}^0\rightarrow \mu^+\mu^-$ decays is its average decay time in an experiment, known as \textit{effective lifetime}. This observable is dependent on the decay width asymmetry between heavy and light $B_{s}^0$ mass eigenstates and the $A_{\Delta \Gamma _s}^{\mu\mu}$ parameter, which is equal to unity in the SM. By measuring the $B_{s}^0\rightarrow \mu^+\mu^-$ effective lifetime, the $A_{\Delta \Gamma _s}^{\mu\mu}$ can be evaluated, and the way each mass eigenstate contributes to the decay can be inferred. The latest LHCb result using full Run 1 and Run 2 data~\cite{prd-bsmumu,prl-bsmumu} yields $\tau_{\mu^+\mu^-} = (2.07 \pm 0.29\pm 0.03)$ ps, which is consistent with the heavy mass eigenstate lifetime, as predicted by the SM, up to 1.5 standard deviations.

\section{Lepton Flavour Non Universality}
In the SM, the electroweak coupling is universal for all the leptons. Differences in the decay rates involving the three lepton species are only expected to arise due to the different masses of the charged leptons. This accidental symmetry of the SM, known as \textit{lepton flavour universality} (LFU), might be violated in BSM scenarios.

Lepton flavour universality can be tested in many different decays by evaluating the ratio between the branching fractions of decays involving different leptons. These ratios are well predicted in the SM since the associated QCD uncertainties largely cancel, making them a good probe for the BSM models. 

\subsection{LFU test in $b\rightarrow sl^+l^-$ transitions}
In the case of the $b\rightarrow sl^+l^-$ transition, the LFU is studied through the branching fraction ratio of decays involving muons and electrons:

\begin{equation}
 R_{H_s} = \frac{\mathcal{B}(X_b\rightarrow H_s\mu ^{+}\mu^{-})}{\mathcal{B}(X_b\rightarrow H_se ^{+}e^{-})}
\end{equation}

\noindent where $X_b$ is a hadron containing a $b$ quark and $H_s$ is a hadron with one $s$ quark.

Although theoretically very clean, the measurement of these ratios is experimentally challenging due to the detection asymmetry between muons and electrons at LHCb. Muons are easy to reconstruct and trigger because they are the only particles reaching the muon chambers, leaving a characteristic signature in the detector. In the case of electrons, the reconstruction and triggering processes are more complex, mainly due to the bremsstrahlung emission undergone by electrons. If the bremsstrahlung takes place after the trajectory of the electron has been deflected by the magnet, the emitted photon will land in the same calorimeter cell as the electron, allowing for a recovery of the photon in the reconstruction of the electron energy. However, if the emission occurs before the magnet, the trajectory of the electron is deflected after the emission, resulting in the photon and the electron to land in different calorimeter cells. In such a case, a procedure to partially recover the energy of the photon is applied, worsening the momentum resolution. 

To reduce the systematics associated with the different detection of muons and electrons, the LFU tests are performed in LHCb using a double-ratio. This approach is used to measure the $R_K$ ratio at LHCb, an analysis performed using the entire Run 1 and Run 2 dataset. The $B^+\rightarrow J/\psi(\rightarrow l^+l^-) K^{+}$ is used as the normalization mode, and the $R_K$ is determined as 

\begin{eqnarray}
 \hspace*{-10cm} R_K & = & \frac{\mathcal{B}(B^+\rightarrow K^+\mu^+\mu^-)}{\mathcal{B}(B^+\rightarrow J/\psi(\mu^+\mu^-)K^+)} \nonumber \\ 
 & & \times \frac{\mathcal{B}(B^+\rightarrow J/\psi(e^+e^-)K^+)}{\mathcal{B}(B^+\rightarrow K^+e^+e^-)}
\end{eqnarray}

The rare mode is selected by requiring $1.1 < q^2_{ll} < 6.0 $ GeV$^2/$c$^4$ to reject contributions from the $J/\psi$ resonant mode and other excited states ($\psi(2S)$ and $\psi(3770)$) at high-$q^2$, and $\phi(1020)$ at low-$q^2$.

\begin{figure}[h]
       \centering
       \includegraphics[width=60mm]{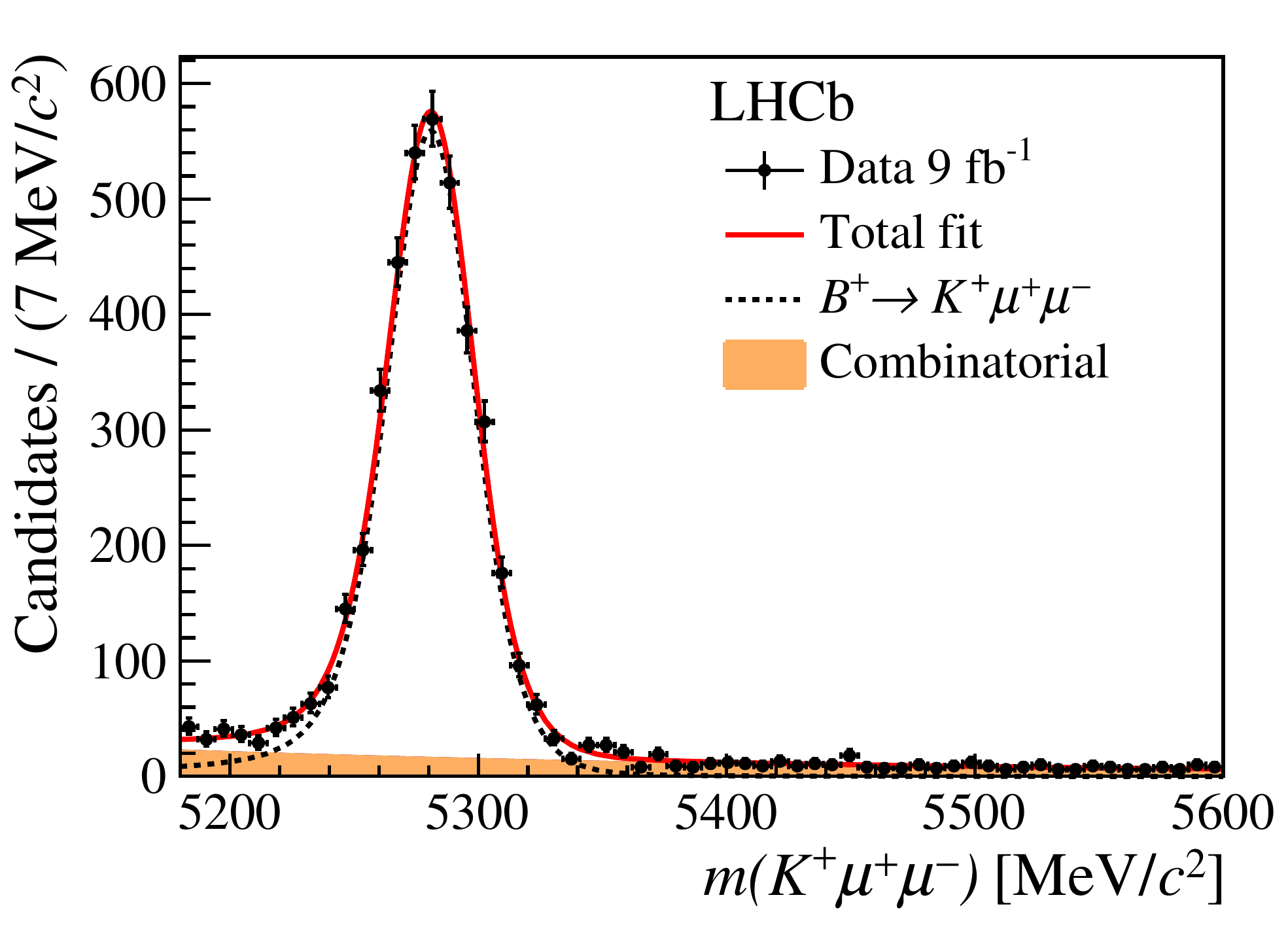}
       \caption{Invariant mass distribution of the selected $B^+\rightarrow K^+\mu^+\mu^+$ candidates. The total fit is overlaid.} 
       \label{fig:rk_muon}
\end{figure}

\begin{figure}[h]
       \centering
       \includegraphics[width=60mm]{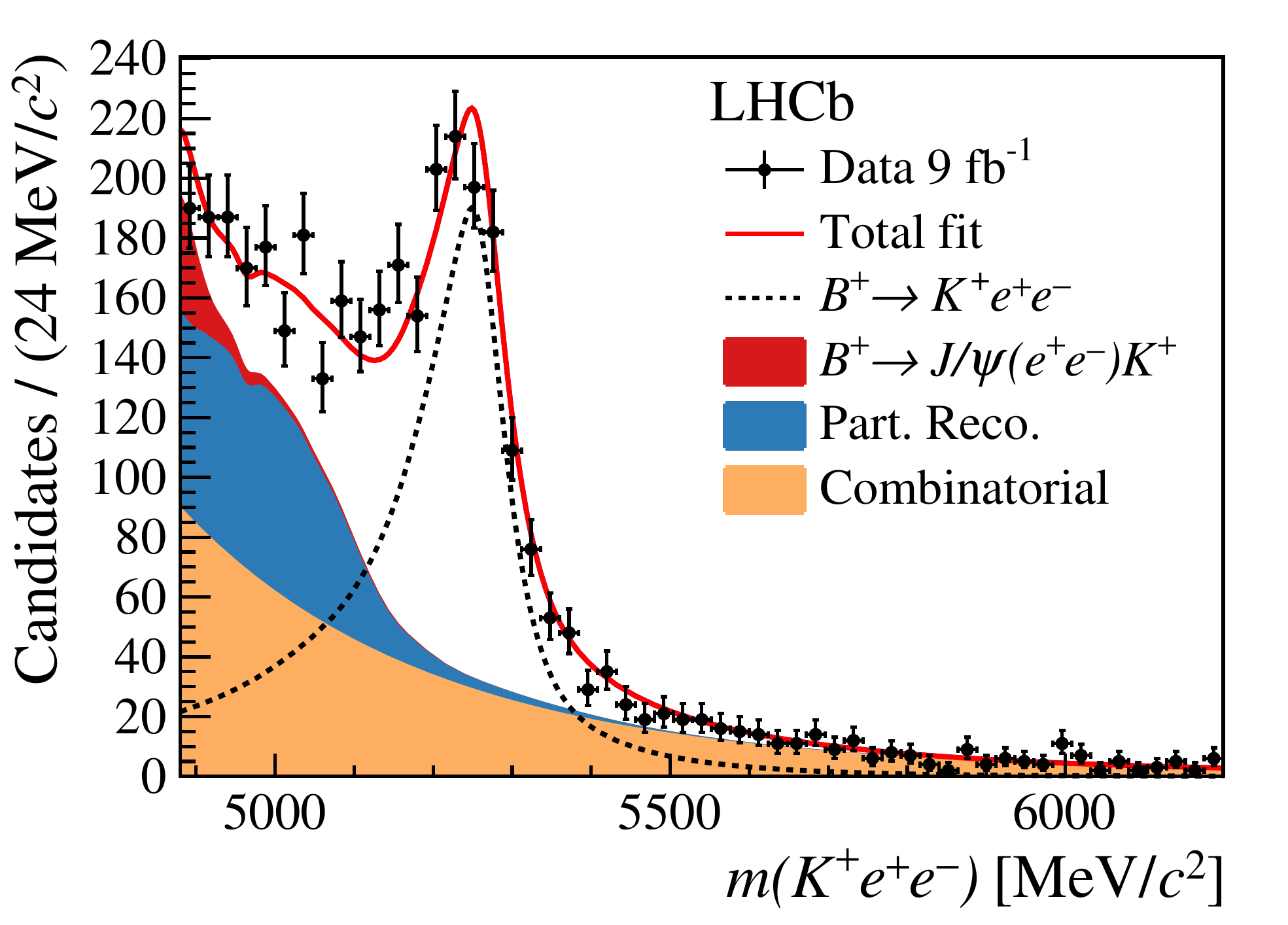}
       \caption{Invariant mass distribution of the selected $B^+\rightarrow K^+e^+e^+$ candidates. The larger background with respect to the muon mode is due to the looser selection in the electron mode.} 
       \label{fig:rk_electron}
\end{figure}

The single ratio $r_{J/\psi} = \mathcal{B}(B^+\rightarrow J/\psi(\mu^+\mu^-)K^+)/\mathcal{B}(B^+\rightarrow J/\psi(e^+e^-)K^+)$ is measured to validate the detection efficiencies. The measured value is consistent with unity as predicted by the SM, and no significant trend is observed in a number of kinematic regions. Since the $r_{J/\psi}$ ratio does not benefit from the cancellation of systematic uncertainties due to the different detection of muons and electrons, this result demonstrates the large control over the relative efficiencies for electrons and muons. The efficiencies and the yields from the resonant modes, obtained from maximum likelihood fits to data samples, are used as input variables to the simultaneous fit of the rare modes, which are shown in Figures~\ref{fig:rk_muon} and~\ref{fig:rk_electron}. The resulting $R_K$ measurement is found to be $R_K = 0.846^{\mbox{ }+0.042\mbox{ }+0.013}_{\mbox{ }-0.039 \mbox{ }-0.012}$, showing a discrepancy of 3.1 standard deviations with respect to the SM~\cite{rk}.

The isospin partner of the $B^+\rightarrow K^+l^+\l^-$, the $B^0\rightarrow K^0_sl^+l^-$ decay, is also measured at LHCb, resulting in the $R_{K^0_s}$ ratio. A similar analysis strategy as for the $R_K$ is followed, using the $B^+\rightarrow J/\psi(\rightarrow l^+l^-) K^{0}_s$ as the normalization channel for the double ratio. The $K_s^0$ is reconstructed as $K_s^0\rightarrow \pi^+\pi^-$, lowering the efficiencies and the precision with respect to the $R_K$ measurement. The result obtained using Run 1 and Run 2 data is $R_{K_s^0} = 0.66^{+0.20+0.02}_{-0.14-0.04}$, found to be 1.5 standard deviations below the SM prediction~\cite{rkshort}.

In addition, the $R_{K^{*+}}$ ratio with $B^+\rightarrow K^{*+}l^+l^-$ as the rare modes has also been measured by LHCb using the full dataset available. The same analysis strategy as for $R_K$ is followed. The $K^{*+}$ is reconstructed as $K^{*+} \rightarrow K_s^0\pi^+$ with $K_s^0\rightarrow\pi^+\pi^-$. A wider range in $q^2$ of $0.045 < q^2_{ll} < 6.0 $ GeV$^2/$c$^4$ is used to include the enhancement of the branching fraction at low $q^2$ caused by the photon pole. A measurement of $R_{K^{*+}} = 0.70^{+0.18+0.03}_{-0.13-0.04}$ is obtained, found to be 1.4 standard deviation below the SM prediction of unity~\cite{rkshort}.

\subsection{LFU test in $b\rightarrow cl\nu$ transitions: $R_{\Lambda_c^+}$}

LFU tests in the baryonic sector offer complementary information to the meson sector due to the spin 1/2 in the initial state. Moreover, the form factor involved in the baryonic transitions are different from the mesonic decays presented so far, probing BSM models in different scenarios. 

The latest measurement on this respect performed by LHCb corresponds to the observation of the $\Lambda_b^0\rightarrow \Lambda_c^+\tau^-\bar{\nu}_{\tau}$ using Run 1 data~\cite{rlambda}. The tau candidates are reconstructed in the decays $\tau^-\rightarrow \pi^-\pi^+\pi^-\nu_{\tau}$ and $\tau^-\rightarrow \pi^-\pi^+\pi^-\pi^0\nu_{\tau}$, from a $\pi^-\pi^+\pi^-$ combination, as the neutral pion is not reconstructed. The $\Lambda_c^+$ candidate is reconstructed as $\Lambda_c^+\rightarrow p K^-\pi^+$. The main source of background, caused by $\Lambda_b^0\rightarrow \Lambda_c^+\pi^-\pi^+\pi^-X$ decays, is reduced by requiring the tau vertex to be downstream and displaced from the $\Lambda_c^+$ vertex. The contributions of double charm processes such as $\Lambda_b^0\rightarrow D_s^0$ are controlled with a Boosted Decision Tree (BDT) which uses the information of the $\pi^-\pi^+\pi^-$ system.

The branching fraction of the $\Lambda_b^0\rightarrow \Lambda_c^+\tau^-\bar{\nu}_{\tau}$ is determined from a 3 dimensional template fit to the pseudo-decay time of the $\tau$, the $q^2_{\tau\nu}$ and the BDT output. From the fit, the branching fraction is found to be $\mathcal{B} (\Lambda_b^0\rightarrow \Lambda_c^+\tau^-\bar{\nu}_{\tau}) = (1.50 \pm 0.16\pm 0.25\pm 0.23)\%$, where the first uncertainty is statistical, the second systematic and the third due to the external branching fraction of $\Lambda_b^0\rightarrow \Lambda_c^+3\pi$, used as the normalization channel. 


LFU can be tested with the $R_{\Lambda_c^+}$ ratio defined as

\begin{equation}
       R_{\Lambda_c^+} = \frac{\mathcal{B}(\Lambda_b^0\rightarrow \Lambda_c^+\tau^-\bar{\nu}_{\tau})}{\mathcal{B}(\Lambda_b^0\rightarrow \Lambda_c^+\mu^-\bar{\nu}_{\mu})}
\end{equation}

Using the known value for $\mathcal{B} (\Lambda_b^0\rightarrow \Lambda_c^+\mu^-\bar{\nu}_{\mu})$~\cite{PDG2020}, $R_{\Lambda_c^+}$ is found to be $R_{\Lambda_c^+} = 0.242\pm0.026 \pm 0.040  \pm 0.059$, where the first uncertainty is statistical, the second systematic and the third due to the external branching fraction. This result is in agreement with the SM prediction of $0.324\pm 0.004$~\cite{sm-rlambda}.

\section{Summary}
Rare processes involving the $b\rightarrow s l^+l^-$ transition are sensitive to New Physics. The latest LHCb measurements of the branching fraction and effective lifetime of the $B_s^0\rightarrow \mu^+\mu^-$ decays agree with SM predictions. Since no significant excess is found for $B_s^0\rightarrow \mu^+\mu^-\gamma$ and $B^0\rightarrow \mu^+\mu^-$, upper limits are set to their branching fractions which are consistent with the SM predictions. 

Lepton flavour universality tests are also an excellent probe to search for New Physics. The latest result obtained by the LHCb for the $R_K$ ratio is three standard deviations below the SM prediction, suggesting a deficit in the muon mode. Although in agreement with the SM, the measurements of the $R_{K_S^0}$ and $R_{K^{*+}}$ also point to the same direction. Studies of the LFU in the baryonic sector provide a complementary check of LFU due to the different form factors involved. The latest result obtained for $R_{\Lambda^{0}_b}$, in agreement with the SM, shows the baryonic sector looks promising to test LFU and search for physics beyond the standard model.

\bigskip 

\end{document}